\documentclass[final]{svjour2}
\usepackage{graphicx}
\usepackage{rotating}
\usepackage{amssymb}
\usepackage{mathptmx}
\usepackage[numbers]{natbib}
\makeatletter
\journalname{Journal of Low Temperature Physics}

\bibpunct{}{}{,}{s}{}{,}
\newcommand{\ignore}[1]{}
\newcommand{\bce}{\begin{center}}
\newcommand{\ece}{\end{center}}
\newcommand{\beq}{\begin{equation}}
\newcommand{\eeq}{\end{equation}}
\newcommand{\beqa}{\begin{eqnarray}}
\newcommand{\eeqa}{\end{eqnarray}}

\begin{document}

\newcommand{\hdblarrow}{H\makebox[0.9ex][l]{$\downdownarrows$}-}
\title{Improved variational approach to the two-site Bose-Hubbard model}

\author{M. Mel\'e-Messeguer$^1$, B. Juli\'a-D\'\i az$^{1,2,3}$,\\ A. Polls$^{1,2}$}

\institute{1: Departament d'Estructura i Constituents de la Mat\`{e}ria,
Universitat de Barcelona, 08028 Barcelona, Spain
\\2: Institut de Ci\`encies del Cosmos,
Universitat de Barcelona, 08028 Barcelona, Spain
\\3: ICFO - Institut de Ci\`encies Fot\`oniques, 08860 Castelldefels, Spain}

\date{04.04.2011}

\maketitle


\begin{abstract}
An improved variational ansatz is proposed to capture 
the most striking properties of the ground state of 
a slightly biased attractive two-site Bose-Hubbard 
Hamiltonian. Our ansatz, albeit its simplicity, is 
found to capture well the exact properties of the 
ground state for a wide variety of model parameters, 
in particular the fragmentation occurring before the formation 
of cat-like states and also the formation of strongly 
correlated cat-like states.

\end{abstract}

\section{Introduction}

The physics of ultracold bosons confined in a 
double-well potential has attracted a great 
deal of attention since the theoretical prediction 
of Josephson-like oscillations of the atoms 
population and the existence of self-trapped 
states~\cite{smerzi97,milburn97}. In addition, 
the recent experimental realization of a bosonic 
Josephson junction (BJJ) by the Heidelberg group 
using $^{87}$Rb atoms~\cite{albiez05} has triggered 
the possibility of practical applications and 
extensions to other physical 
scenarios~\cite{ashhab02,lewenstein-adv,julia09,julia10,julia10-3,marta11}. 

The theoretical prediction of Smerzi {\it et. al.}~\cite{smerzi97} was 
made by means of the mean-field Gross-Pitaevskii 
(GP) equation~\cite{pethick2002,pitaevski2003,leggett01}, 
which correctly captures the tunneling dynamics of 
the population and its coupling to the phase 
difference between the two sides of the barrier. 
A further simplification, which turned out to be 
particularly useful, is the consideration of only 
the lowest two modes of the GP equation. Most of 
the semi-classical predictions of this 
two-mode approach~\cite{Raghavan99,maya,julia10}, 
dealing with the Rabi to Josephson transition, have 
been confirmed in a Josephson experiment where the two 
modes are two distinct internal states of the atom~\cite{tilman10}. 

The two-mode model can be requantized giving 
rise to a two-site Bose-Hubbard (BH) 
model~\cite{milburn97,leggett01,bloch-rmp}. It is worth noting that 
the regime of applicability of the quantized two 
mode approximation can extend further: recent examples 
are the experiments on BJJ, the production of number 
squeezed states, and a non-linear atom 
interferometer~\cite{esteve08,gross10}. These phenomena 
are beyond GP, as they involve entangled states of the 
atoms in the cloud, but can, however, be explained 
within the Bose-Hubbard model~\cite{milburn97,cirac98}. 

The two site BH model predicts for the case of 
attractive interactions the existence of a strongly 
correlated ground state for specific values of the 
parameters~\cite{cirac98,julia10}. These ground 
states are far from being of mean-field type 
thus exhibiting interesting quantum properties, e.g. 
cat-state-like behavior. The existence of strongly 
correlated ground states of quantum systems has recently 
been linked to the existence of instabilities of the 
semi-classical predictions in several different 
contexts: sonic analogues of black holes~\cite{garay}, 
vortex nucleation in small atomic 
clouds~\cite{ueda,dagnino09,wilkin}, Bose-Einstein condensates (BEC) 
in rotating ring superlattices~\cite{nunnenkamp}, 
or in the ground state of BEC in a double-well potential~\cite{weiss}. 

The simplicity of the two-site BH model allows for an exact 
numerical solution. However, it is always useful to have analytical 
insight which captures the essential physics sometimes hidden 
in the numerical diagonalization process of the Hamiltonian. 
In Ref.~[\citenum{cirac98}], a mean-field state was proposed 
and the need to go beyond this approach was clearly established 
especially in the bifurcation region where the cat-like states 
were identified. Later works along the same lines, but more 
focused on the dynamical properties, have contributed 
to the analysis of these systems~\cite{holthaus01}. 
Based on the limitations of 
the existing variational states, in this manuscript 
we propose an improved variational ansatz which is shown to 
yield an accurate description of the exact ground state of the 
system for a broad range of interaction strengths, including the strongly 
correlated regimes. This improved  trial state is 
constructed by combining two states of 
mean-field type, thus also providing an analytical 
representation of the ground state of the system, 
capturing its most representative features. 

The manuscript is organized as follows. In Sect.~\ref{sec1} 
we recall the definition of the two-site Bose-Hubbard 
model and introduce the tools 
to analyze the system. We also comment on some 
standard results obtained by exact diagonalization. 
Following the steps of Ref.~[\citenum{cirac98}], 
in Sect.~\ref{sec2} we analyze  the advantages and 
limitations of the mean-field approximation, and 
explore the possibilities of a variational state 
that describes the system beyond mean-field.
In Sect.~\ref{crit} we propose an improved variational 
state that can be used in the full range of 
the interaction strength and that incorporates the 
mean-field description, the incipient fragmentation before
the bifurcation and the strongly correlated 
cat-like states whenever they are present. 

The main 
results and the conclusions are summarized in the last section.

\section{Theoretical description}
\label{sec1}

A good description of a system with $N$ particles that 
populate two weakly coupled states, which could represent 
the two sides (left and right) of a double-well, and with 
weak interaction between the particles that occupy the same 
state, is provided by the Bose-Hubbard model,  
\begin{eqnarray}
H &=& 
-\varepsilon (a_L^{\dagger} a_L - a_R^{\dagger} a_R) 
- J (a_L^{\dagger}a_R + a_L a_R^{\dagger}) + 
\frac{U}{2} (
a_L^{\dagger} a_L^{\dagger} a_L a_L +
a_R^{\dagger} a_R^{\dagger} a_R a_R 
)\,. \label{eq:ham}
\end{eqnarray}
Where $J$ describes the coupling between the two states, 
i.e. tunneling in the case of a double-well. Here, 
$U$ characterizes the interaction between the particles 
and is taken to be the same in both sites. $U>0$ 
($U<0$) describes a repulsive (attractive) interaction. 
A small bias, $0 < \varepsilon \ll J$, is introduced to 
ensure the breaking of the left-right symmetry. 
Positive values of $\varepsilon$ promote the $L$ 
state~\footnote{Note that from here on in our discussion,  
we will use the nomenclature of two sites or two 
wells when we refer to the two weakly coupled states 
that define our Bose-Hubbard model.}.

A natural basis to study the system is the Fock basis, 
which is characterized by the number of atoms in each 
of the two modes, $| N_L,N_R \rangle$. 
This basis, $\left \{ | N,0 \rangle, | N-1, 1 \rangle , ..., | 1, N-1 \rangle ,
| 0,N \rangle \right \}$ spans an $N+1$ dimensional 
space, where $N=N_L+N_R$ is the total number of particles. 

The action of the creation and annihilation operators 
on these states is defined in the following way: 
$a_L^{\dagger} | N_L,N_R \rangle =\sqrt{N_L+1} | N_L+1, N_R \rangle$, 
and $a_L | N_L,N_R \rangle = \sqrt {N_L} | N_L-1,N_R \rangle$. 
Therefore, 
\begin{equation}
| N_L,N_R \rangle = \frac {1}{{\sqrt {N_L! N_R!}}} 
(a_L^{\dagger})^{N_L} (a_R^{\dagger})^{N_R} | 0,0 \rangle \,  . 
\end{equation}
In the two-mode approximation, a general N-body state can be written as 
\beq
|\Psi \rangle = \sum_{k=0}^N c_k | k, N-k \rangle \,,
\eeq
and the average number of atoms in each mode for a given 
state is 
$N_{\beta} = \langle \Psi | a_{\beta}^{\dagger} a_{\beta} | \Psi\rangle $, 
with $\beta = L,R$. The population imbalance of a state 
$|\Psi\rangle$ and its dispersion are defined as:
\beqa
z = \langle \Psi | \hat{Z} | \Psi\rangle \; ; \qquad  
\sigma_z = {\sqrt {\langle  \Psi | \hat{Z}^2 | \Psi \rangle - \langle \Psi |  \hat{Z} | \Psi \rangle^2 }}
\label{eq:imba-disp} \;.
\eeqa
with $\hat{Z}= ( a_{L}^{\dagger} a_{L} - a_{R}^{\dagger} a_{R})/N$. 

To characterize the degree of condensation of the system
one can make use of the one-body density matrix, $\rho$. 
For a state $|\Psi\rangle$, we have 
$\rho_{ij} = \langle \Psi | \hat \rho_{ij} | \Psi\rangle$, with 
$\hat \rho_{ij}= a_i^{\dagger}a_j$ and $i,j=L,R$. The trace 
of $\rho$ is normalized to the total number of atoms. The 
two normalized eigenvalues, i.e. eigenvalues divided by the total number 
of atoms $N$, of $\rho$ are $n_{1(2)}$, with $n_1>n_2$. They 
fulfill $n_1+n_2=1$. The eigenvalue $n_i$ corresponds to the condensate 
fraction in the macro-occupied single-particle state 
$|\psi_i\rangle$ which is the $i$-th eigenvector of the one-body density 
matrix. When the eigenvalues of the density matrix are 
strictly $n_1=1$ and $n_2=0$, the system is fully condensed 
in a single-particle state $\psi_1$ (eigenvector of $\rho$). 
In this case, it is possible to express $|\Psi\rangle$ with 
a mean-field state constructed as 
$| \Psi \rangle_N = | \psi_1 \rangle\otimes \dots 
\otimes  | \psi_1 \rangle \equiv | \psi_1\rangle ^{\otimes N}$.

\subsection{Exact spectral properties}

In this work we fix $J=1$, which is equivalent to measuring  
the energy in units of $J$. We will vary the number of 
particles $N$ and the strength of the interaction, which 
will be considered always attractive, $U < 0$. The bias 
term, that can be related to possible small asymmetries 
of the external potential, will be taken very small: 
$\varepsilon /J = 10^{-8}$.

\begin{figure}[t]
\includegraphics[width=0.51\columnwidth, clip=true]{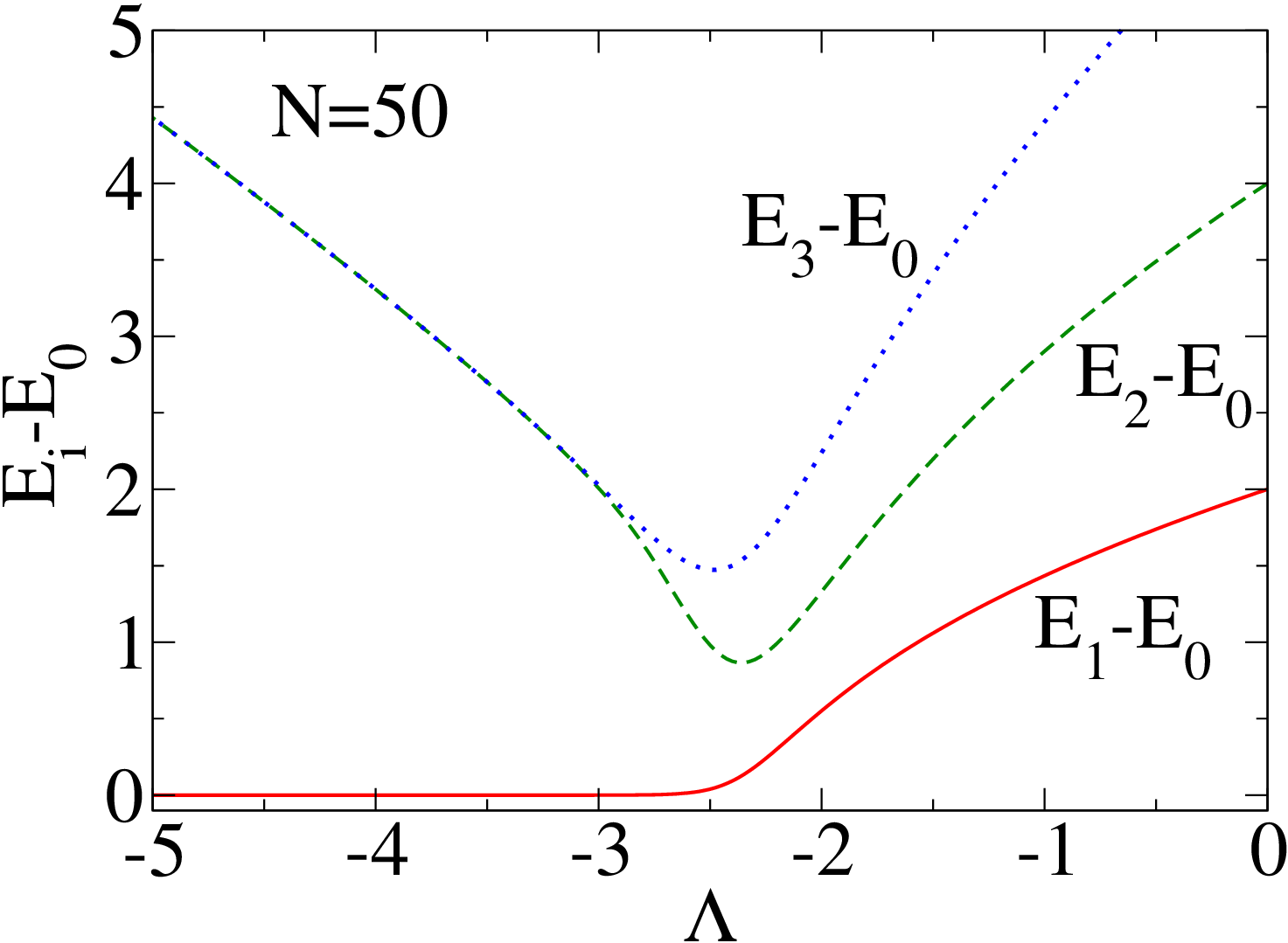}
\includegraphics[width=0.47\columnwidth, clip=true]{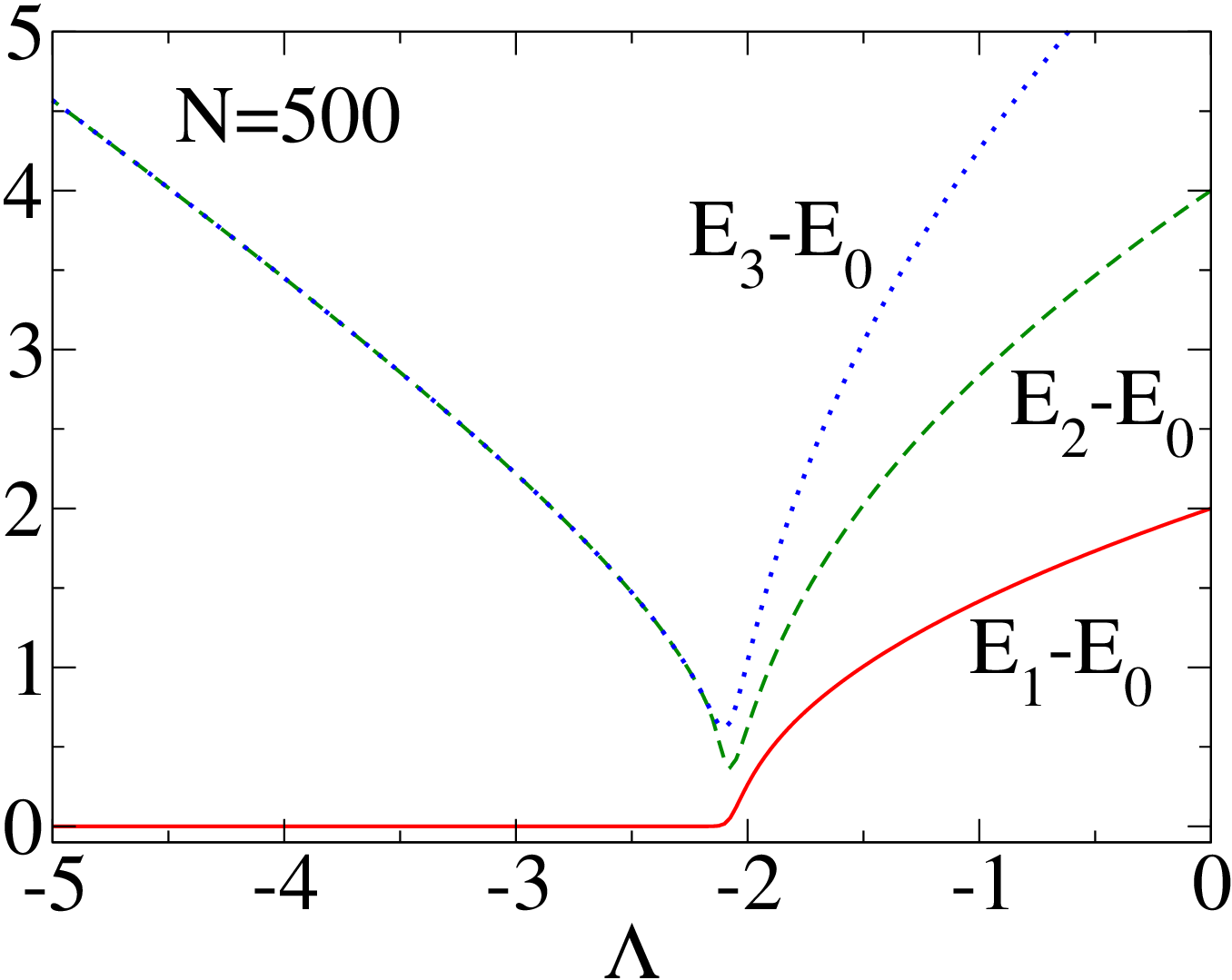}
\caption[]{(Color online) Energies of the lowest energy 
levels with respect to the ground state energy $E_0$, 
as a function of the parameter $\Lambda=NU/J$, for $N=50$ 
(left panel) and $N=500$ (right panel). All energies are 
measured in units of $J$.} \label{fig:fig1}
\end{figure} 

In principle, one can calculate the matrix elements 
of the Hamiltonian in the Fock-space basis and by 
diagonalization obtain the full spectrum of the system 
and the spectral decomposition of the eigenstates~\cite{cirac98,holthaus01,julia10,meystre,trippenbach}.
It is then straightforward to calculate also the population 
imbalance and the degree of condensation of each state.

In this paper we are interested in understanding the 
physical nature of the ground state of the system, 
which is for some parameter values quasi-degenerate with 
the first excited state. Therefore, we start by considering 
the lowest energy levels of the system as a function of 
$\Lambda \equiv NU/J$, a parameter governing the behavior 
of the system. In Fig.~\ref{fig:fig1}, we report the 
energies of the first three excited states with respect 
to the ground state of the system as a function of 
$\Lambda$ for two different numbers of particles, 
$N=50$ and $N=500$, obtained by direct 
diagonalization~\cite{cirac98}. For vanishing atom-atom interaction, 
$\Lambda =0$, the energy gap for consecutive states 
is equal (except for the bias), and the gap is 
independent of the number of particles. As $|\Lambda|$ 
increases the eigenvalues start to merge in pairs (the 
ground with the first excited, the second with the 
third, etc.) but due to both $\varepsilon$ and $J$, 
they do not reach complete degeneracy. Moreover, the 
convergence of the merging process depends on the number 
of particles: for higher $N$ it occurs at smaller values 
of $|\Lambda |$ reaching the value  $|\Lambda| =2$, when 
the number of particles tends to infinity. 

\begin{figure}[t]
\includegraphics[width=0.51\columnwidth,height=0.25\textheight,clip=true]
{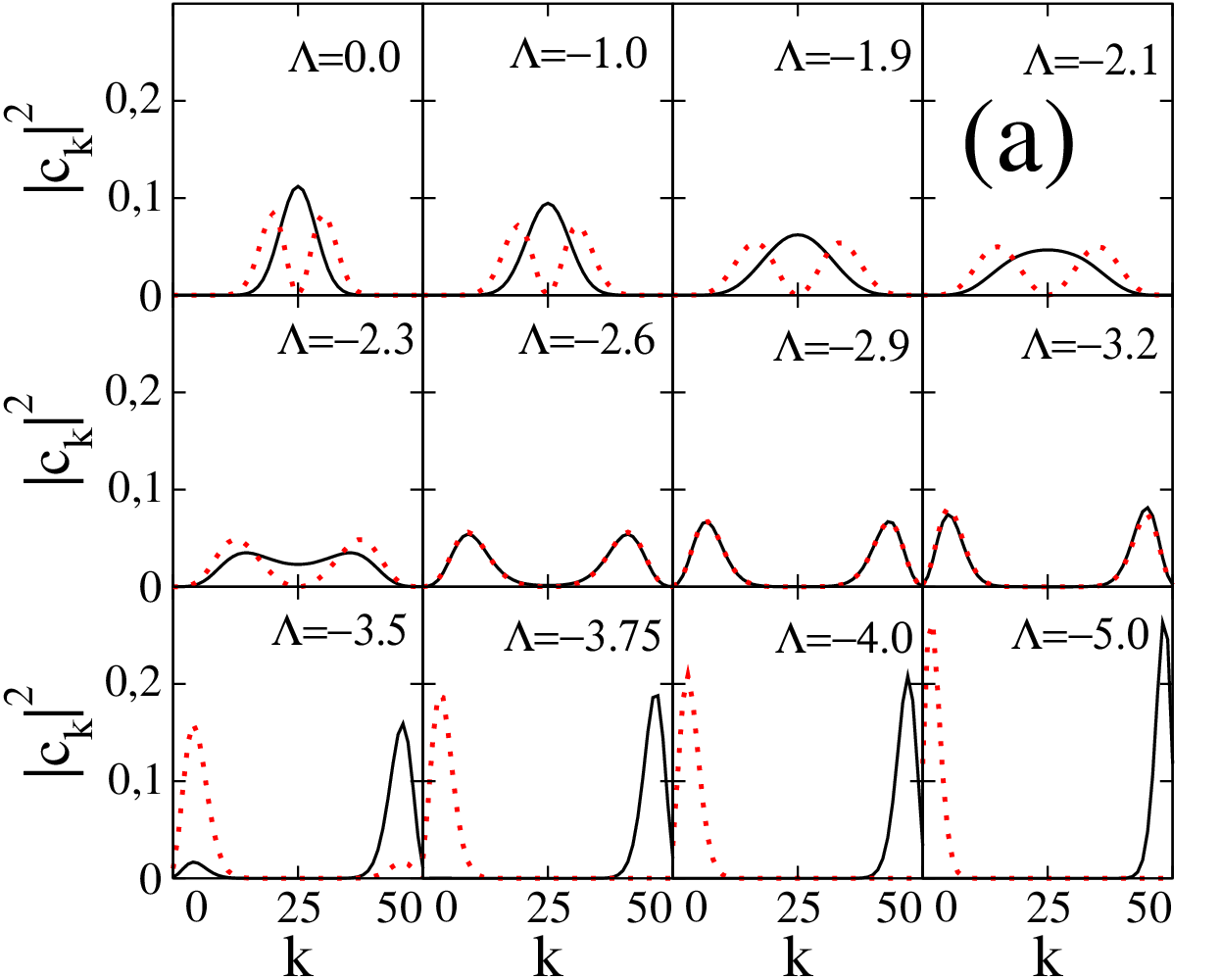}
\includegraphics[width=0.47\columnwidth,height=0.25\textheight,clip=true]
{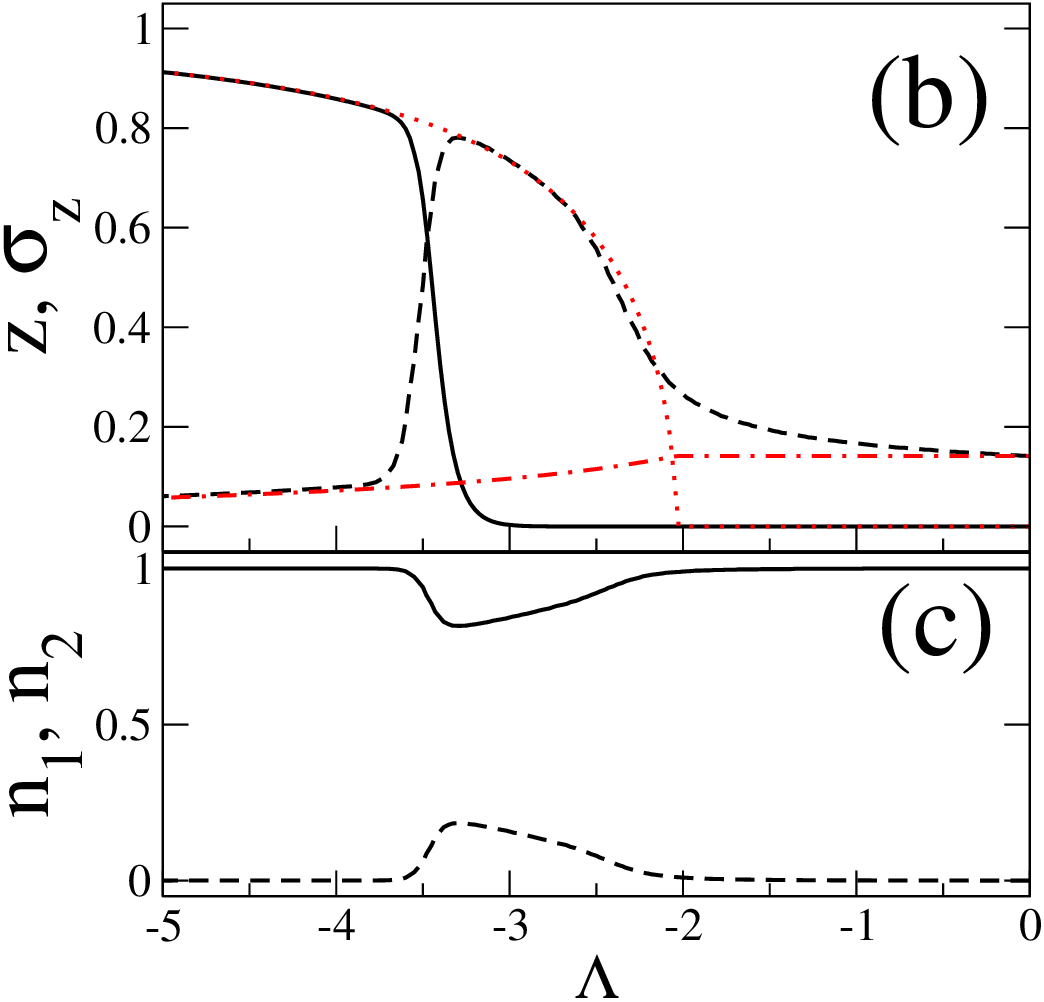}
\caption[]{(Color online) (a) Spectral decomposition 
($| c_k |^2$) in the Fock space of the ground 
(black solid  line) and first excited (red dotted line) states 
for different values of $\Lambda$, with $N=50$.
To help in the reading of the figure, instead of plotting 
the discretized values $|c_k|^2$ we have generated a 
smooth curve by joining the different points. 
(b) Population imbalance $z$ 
(black solid line) and its dispersion $\sigma_z$, Eqs.~(\ref{eq:imba-disp}),
(black dashed line) 
of the exact ground state as a function of $\Lambda$. The semi-classical 
predictions of the imbalance (red dotted line) and its dispersion 
(red dot-dashed) are also plotted~\cite{julia10,julia2-10}.
(c) Solid and dotted lines depict the condensed fractions 
$n_1$ and $n_2$ of the one-body density matrix of the 
exact ground state as a function of $\Lambda$. In all cases 
$N=50$.
} 
\label{fig:fig2}
\end{figure}

In Fig.~\ref{fig:fig2} (a) we plot the spectral decomposition 
of the ground and first excited states in the Fock space 
for different values of $\Lambda$, and for $N=50$. The 
plotted values $|c_k|^2$ give the probability that the 
state has $k$ particles in the left well and $N-k$ 
particles in the right one. Notice that if the spectral 
decomposition of the state is peaked at high values of 
$k$, it means that for this state most of the atoms 
are located on the left side of the double-well. For 
weak interactions, $|\Lambda|< 2.6$, the spectral 
decomposition of the ground and the first excited states 
are clearly different, (as were also the energies in 
Fig.~\ref{fig:fig1}). For stronger interactions, 
$-3.2 \le \Lambda \le -2.6$, the two states 
become very close in energy (Fig.~\ref{fig:fig1}, 
left panel) and their spectral decompositions $|c_k|^2$ 
are very similar. However, one should notice that 
the ground state is symmetric, 
$c_{N/2+k} = c_{N/2-k}$, and the first excited one is 
antisymmetric, $c_{N/2+k} = -c_{N/2-k}$. In this region 
the ground state is a strongly correlated cat-like 
state~\footnote{Strictly speaking, the purest cat-state 
would correspond to the state 
${1/\sqrt{2}}(|N,0\rangle + |0,N\rangle)$. The states 
we refer to as cat-like states are sometimes called kitten states with a 
certain degree of 'catness'~\cite{ciracats}.} as its 
spectral decomposition has two clear peaks.

Finally, for $|\Lambda| > 3.2 $, the two states become 
again clearly different: the ground state is peaked at 
a high value of $k$, with a large amount of atoms in 
the left well, while the first excited has its peak 
at a low value of $k$. Note that the energies of 
these states are very close to each other.
  
A useful characterization of the ground state is provided 
by the population imbalance $z$. As shown in Fig.~\ref{fig:fig2} (b), 
it remains zero up to a certain value of $|\Lambda|$ 
($\sim 3.25$ for $N=50$), approaches 1 as $|\Lambda|$ 
increases further. The figure also shows $\sigma_z$, which starts 
from small values associated to a relatively narrow 
binomial distribution. It increases in the range where 
the strongly cat-like state is present, and finally 
decreases abruptly when $|\Lambda|$ increases further 
and the ground state populates massively the $L$ state. 
Thus $z \rightarrow 1$ and $\sigma_z \rightarrow 0$ for $|\Lambda| \rightarrow
\infty$. 

The degree of condensation of the ground state, 
$|\phi_{\rm gs}\rangle$, is 
determined by the eigenvalues $n_1$ and $n_2$ of 
the one-body density matrix, which are plotted in  
Fig.~\ref{fig:fig2} (c). These condensate fractions 
measure the macroscopic occupations of the single-particle 
states $| \psi_1 \rangle$ and $| \psi_2\rangle$, eigenfunctions 
of $\langle\phi_{\rm gs}|\hat \rho |\phi_{\rm gs}\rangle$. 
The regions where these values are 
not close to 1 and 0, signal the occurrence of fragmentation 
of the ground state and the impossibility to describe the system by means 
of a mean-field state.  
In the region, $-2 <\Lambda < 0$, $n_1$ is rather close to 
1 ($n_1 \sim  0.99$), and the macroscopically occupied state is given by
$| \psi_1\rangle = (| L \rangle + | R \rangle)/{\sqrt 2}$, with 
$|L (R)\rangle \equiv a^\dagger_{L(R)}|0\rangle$. However, as 
we will discuss later in Fig.~\ref{fig:fig3}, this slight fragmentation 
produces noticeable differences in the spectral decomposition of the 
mean-field state build with the state $|\psi_1\rangle$ and the 
exact ground state. The fragmentation is particularly 
important for $-2.5>\Lambda>-3.5$, which is roughly the same interval
where the cat-like structure takes place. However, the macro-occupied
state $|\psi_1\rangle$ remains equal to the one previously discussed. 
The correlations beyond mean-field affect the degree of condensation, 
but not the single state that is mainly occupied. This is because the ground 
state remains almost symmetric (except for the bias) in the Fock space
(see Fig.~\ref{fig:fig2} (a) and \ref{fig:fig3}), i.e. with $z$ almost 
zero. This is reflected in the symmetric character 
of the one-body density matrix, which in turn implies that $|\psi_1\rangle$ 
is the normalized symmetric combination of $|L\rangle$ and $|R\rangle$.
For further increasing $|\Lambda|$, the system becomes again 
condensed: $n_1 \rightarrow 1$. The slight energy difference 
introduced by the bias term, which energetically promotes 
the $|L\rangle$ over the $| R\rangle$ state, 
drives the system to $| \psi_1 \rangle \rightarrow | L \rangle$. 

The precise value of the bias term has been shown to determine, 
for a fixed $N$, the size of the cat-like-region. Exploring the 
interplay between the bias term and the hopping strength, $J$, 
a good estimate of the precise value of $\Lambda$ where the 
bias dominates is given in Ref.~[\citenum{julia2-10}]. 
For larger number of particles, $N$, the bias term becomes 
dominant at lower values of $|\Lambda|$~\cite{julia2-10}, as its effect 
is proportional to $N$, and therefore the cat-like-region becomes 
narrower at values of $\Lambda$ closer to the critical classical 
value $\Lambda = 2$.

\section{Variational state for the ground state}
\label{sec2}

\subsection{Mean-field ansatz}
\label{sec:bmf}

A reliable mean-field state~\cite{cirac98} can be constructed 
using a general single-particle state 
$| \phi \rangle_{\rm sp}= \alpha | L \rangle  + \beta | R \rangle $, 
with $ | \alpha | ^2 + | \beta  |^2 =1 $, and considering 
all the particles to be in this single-particle state:
\begin{equation}
|\phi \rangle_N = \frac {1}{{\sqrt {N!}}} \left [ \alpha a_L^{\dagger} + 
\beta a_R^{\dagger} \right ]^N | 0 \rangle \,.
\end{equation}
The expectation value of the Hamiltonian for this state is, 
\beqa
E(\alpha, \alpha^*,\beta, \beta^*) = \langle \phi | H | \phi \rangle_N
&=& - \varepsilon N(\alpha \alpha^*
- \beta \beta^*) - JN (\alpha^* \beta + \alpha \beta^*) \nonumber \\
&& +\frac {U}{2} N (N-1) ( | \alpha|^4 + | \beta |^4 ) \, .
\eeqa
The  minimization of the energy with respect to the 
variational parameters, together with the normalization 
condition $| \alpha|^2+ | \beta |^2 =1$, yields the 
following equation
\begin{equation}
2 \varepsilon N - J N \left ( \frac {\alpha^2 - \beta ^2 }{\alpha \beta} 
\right ) + U N(N-1) ( | \beta|^2 - | \alpha |^2 ) = 0\,.
\end {equation}
The possible solutions of the previous equation 
will be of the type $(\alpha, \pm \beta)$ with both 
$\alpha$ and $\beta$ positive real numbers. Explicit 
simple analytic solutions to the previous equation can 
be obtained by neglecting the bias term. Therefore, 
taking $\varepsilon =0 $ and introducing 
$\tilde \Lambda  = \Lambda (N-1)/N$, one gets the 
following set of solutions~\cite{cirac98}:
\begin{equation}
\alpha_0 = \beta_0 = \frac {1}{{\sqrt{2}}}\,,
\quad \alpha_{\pm} = \beta_{\mp} = 
{\sqrt {\frac {1}{2} \pm {\sqrt {\frac {1}{2^2}- \frac {1}{\tilde
          \Lambda^2}}}}} \,  ,
\label{eq:alphas}
\end{equation}
that give rise to the multi-particle states:
\begin{equation}
| \phi_i^{\pm} \rangle _N = \frac {1}{{\sqrt {N!}}} 
\left [ \alpha_i a_L^{\dagger} \pm \beta_i a_R^{\dagger} \right ]^N | 0 \rangle   \, ,
\end{equation}
with $i=0,+,-$. Note that the solutions $\alpha_{\pm}$ and 
$\beta_{\pm}$ only exist when $|\tilde \Lambda | > 2$. The 
expectation value of the energy in these states, with 
$\varepsilon =0$, is :
\begin{equation}
E_0^{\pm} \equiv \langle \phi_0^{\pm} | H |\phi_0^{\pm}\rangle=
 \frac {U}{4} N(N-1) \mp JN \, ,
\end{equation}
\begin {equation}
E_+^{\pm}\equiv \langle \phi_+^{\pm} | H |\phi_+^{\pm}\rangle
 = E_-^{\pm} \equiv \langle \phi_-^{\pm} | H |\phi_-^{\pm}\rangle
= \frac {U}{2} N(N-1) 
- \frac {NJ}{\tilde \Lambda} ( 1 \mp 2) \,.
\end{equation}
As $U<0$ and $J=1$, the states $|\phi_i^+\rangle_N$ have 
a lower mean energy than the $| \phi_i^-\rangle_N$ in all 
cases. To find the lowest energy, we study the difference 
between $E_0^+$ and $E_+^+$ (notice that $E_+^+=E_-^+$),
\begin{equation}
E_0^+ - E_+^+ = - NJ \left [ \frac {1}{4} \tilde \Lambda 
+ \left ( \frac {1}{\tilde \Lambda} + 1 \right ) \right ]\,.
\end{equation}
Thus for $|\tilde \Lambda | < 2$, 
the lowest energy state is $| \phi_0^+ \rangle_N$, and for 
$\tilde \Lambda < -2$, both states $| \phi_+^+ \rangle_N$ and 
$ | \phi_-^+ \rangle_N $ have the same minimum mean energy. 

\subsection{Variational ansatz beyond mean-field }

In the case that $E_{\pm}^+$ are the smallest mean-field 
energies, i.e., for $|\tilde{\Lambda}| > 2$, one 
can propose an alternative ansatz~\cite{cirac98} for 
the multi-particle many-body state that goes beyond the 
mean-field approach and tries to incorporate the 
cat-like structure~\footnote{Note the only difference with 
the state used in Ref.~[\citenum{cirac98}] is due to our state 
being normalized to 1.}:
\begin{equation}
| \phi_{\rm cat}\rangle_N = 
{1\over \sqrt{2}}\sqrt{ {1\over 1+\big({2/ |\tilde{\Lambda}|}\big)^N}}
\Big( | \phi_+^+ \rangle_N 
+ | \phi_-^+ \rangle _N  \Big).
\label{eq:cat}
\end{equation}
The expectation value of the Hamiltonian for this 
many-body state,
\begin{equation}
\langle \phi_{\rm cat} | H| \phi_{\rm cat} \rangle_N 
= {NJ\over 4\tilde{\Lambda}}
\Bigg[{4-\tilde{\Lambda}^2\over 1 +(2/|\tilde{\Lambda}|)^N}
+3\tilde{\Lambda}^2 \Bigg]\,,
\end{equation}
is smaller than $E_{\pm}^+$. This state is a 
linear combination of two non-orthogonal mean-field 
states having the same energy expectation 
value (if the bias is not taken into account), but 
two different spectral decompositions. It is precisely 
the fact that they are not orthogonal that allows 
the mean energy value in the state $| \phi_{\rm cat} \rangle$
to be smaller than  $E_+^+ = E_-^+$.

\begin{figure}[t]
\includegraphics[width=0.9\columnwidth,clip=true]{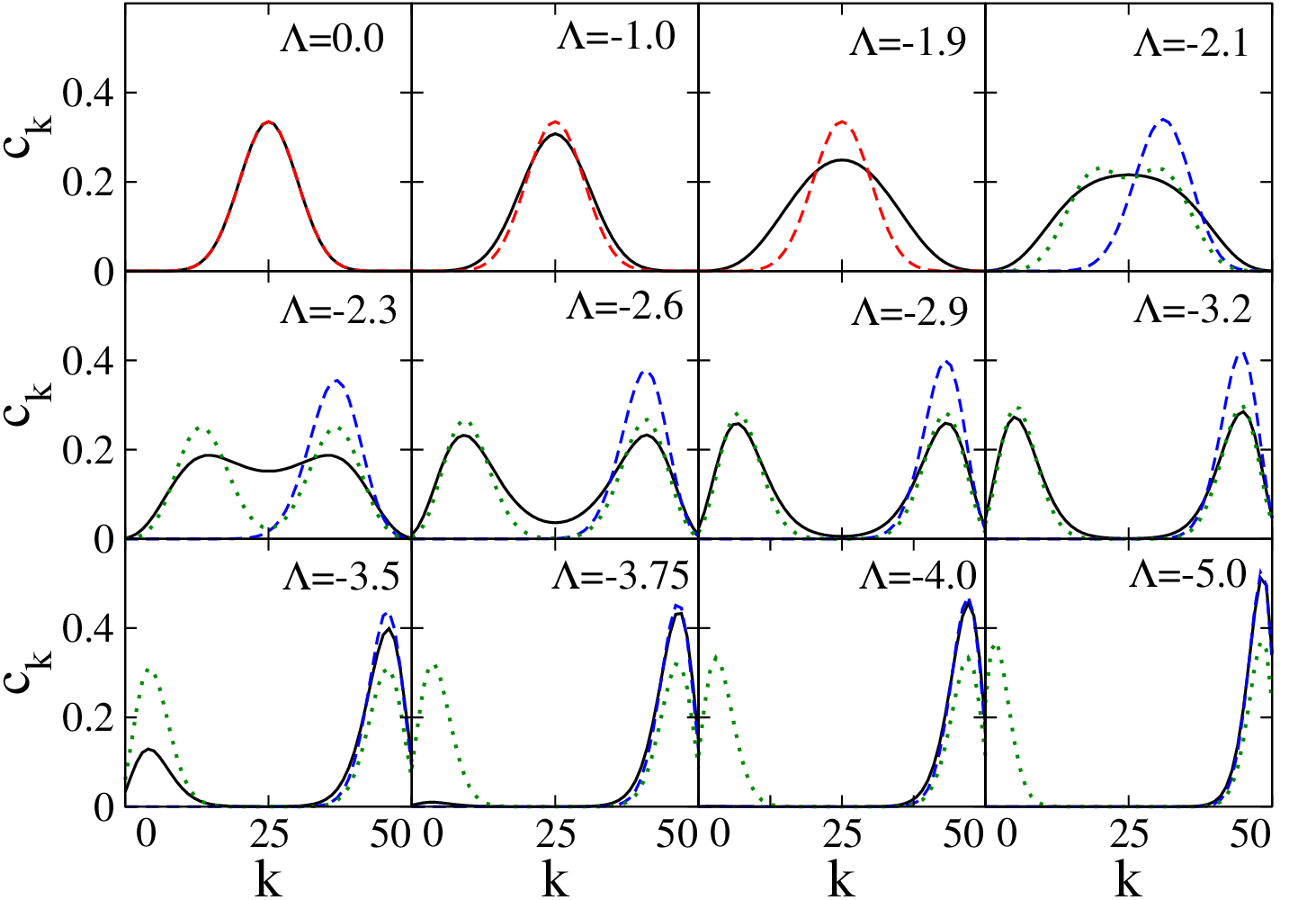}
\caption[]{(Color online) Spectral decomposition ($c_k$) 
in Fock space of the ground state of the system, for 
different values of $\Lambda$, computed by exact diagonalization 
(black solid line) and compared to the spectral decomposition 
of the mean-field functions $| \phi_0^+ \rangle _N$ 
for $| \Lambda | < 2$ (red dashed line) and 
$| \phi_+^+\rangle_N$ for $| \Lambda | >2$ 
(blue dashed line). For $|\Lambda| > 2$,  we 
also show the results for the variational cat-state 
$| \phi_{\rm cat}\rangle$ (green dotted line). 
In all cases $N=50$. 
\label{fig:fig3}}
\end{figure}

In Fig.~\ref{fig:fig3}, we show the Fock space decomposition 
($c_k$) for different values of $\tilde \Lambda$ of 
the ground state of the system computed by exact diagonalization 
of the many-body Hamiltonian, Eq.~(\ref{eq:ham}). We compare these 
coefficients with the ones provided by the mean-field state
$| \phi_0^+\rangle_N$ for $| \tilde \Lambda | < 2$, 
and $| \phi_+^+ \rangle_N$ for $| \tilde \Lambda | > 2$. 
In this last case we also plot the results for the variational 
state $| \phi_{\rm cat} \rangle_N$ given in Eq.~(\ref{eq:cat}). In 
all cases $N=50$. Note that for this number of particles 
$\Lambda$ and $\tilde \Lambda$ are very similar  
and therefore the critical value of $\Lambda$ where the 
mean-field states $|\phi_{\pm}^+ \rangle_N$ 
appear is $\Lambda \sim \tilde \Lambda =-2$.

\begin{figure}[t]
\includegraphics[width=0.49\columnwidth,  clip=true]{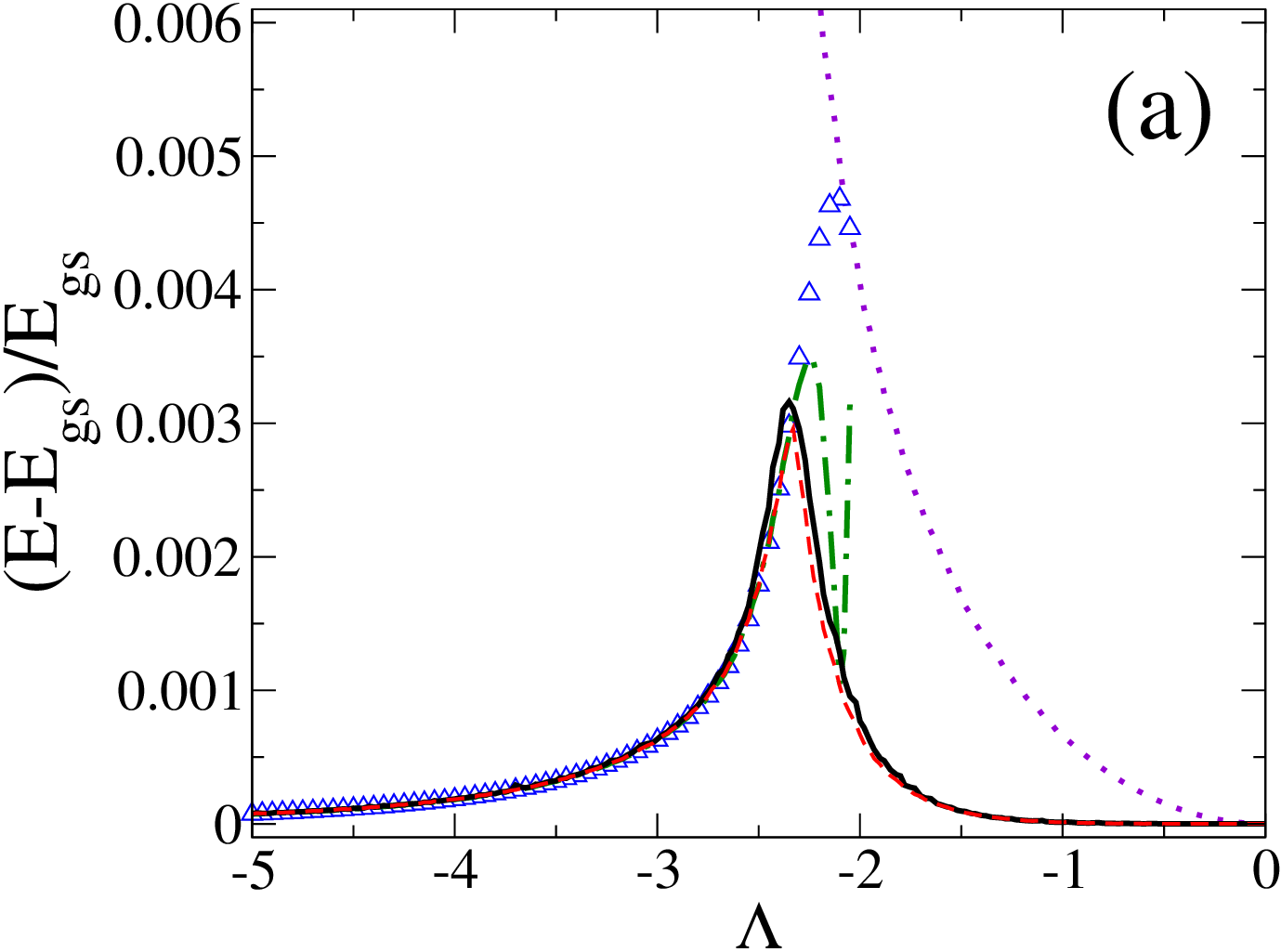}
\includegraphics[width=0.47\columnwidth,  clip=true]{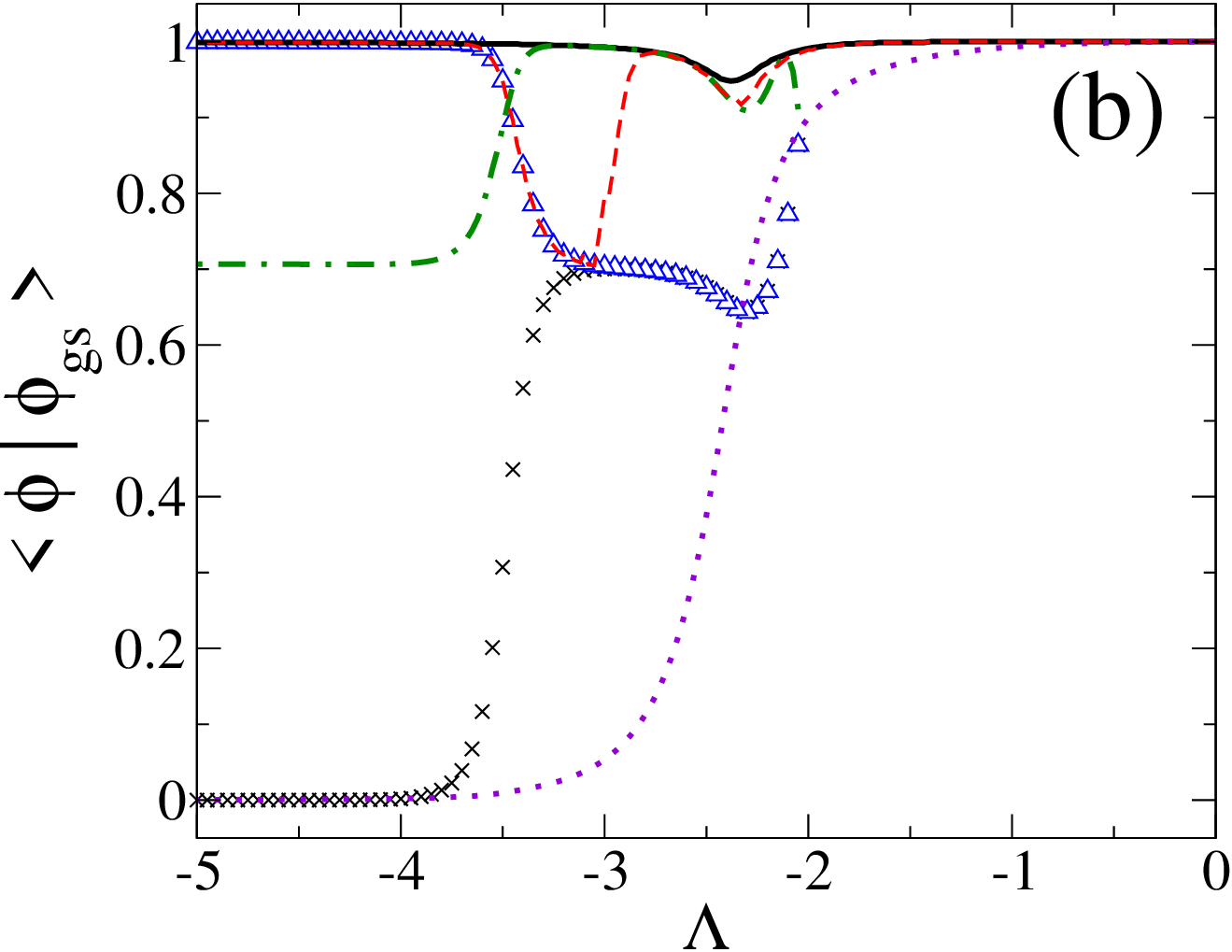}
\caption[]{(Color online) (a) Relative difference with respect to the 
exact ground state energy of the expectation value of the 
Hamiltonian with different variational many-body states:
$| \phi_0^+ \rangle$ (violet dotted line),
$| \phi_{\pm}^+ \rangle$ (blue triangles),
$| \phi_{\rm cat}\rangle$ (Eq.~\ref{eq:cat}) (green dot-dashed line), 
$| \Psi_{\rm var}\rangle_{\rm min}$ (red dashed line) and 
$| \Psi_{\rm var} \rangle_{\rm max}$, (black solid line).
(b) Overlap of the different states discussed 
in the text with the exact ground state of the system as a 
function of $\Lambda$. 
$| \phi_0^+ \rangle$ (violet dotted line), 
$| \phi_{+}^+ \rangle$ (blue triangles), 
$| \phi_{-}^+ \rangle$(black crosses), 
$| \phi_{\rm cat}\rangle $ (Eq.~\ref{eq:cat}) (green dot-dashed line), 
$| \Psi_{\rm var}\rangle_{\rm min}$ (red dashed line) and 
$| \Psi_{\rm var} \rangle_{\rm max}$ (black solid line).}   
\label{fig:fig4} 
\end{figure}

For $| \tilde \Lambda | < 2$, the best mean-field 
representation of the ground state corresponds to 
$| \phi_0^+\rangle_N$. The coefficients $c_k$ follow a 
binomial distribution, symmetric around $k=N/2$. This 
mean-field state gives a good qualitative 
description of the system in this range, however it 
coincides with the exact solution only for $\tilde \Lambda =0$. 
In Fig.~\ref{fig:fig3} one can appreciate that the 
distribution of the exact ground state is slightly broader 
and the differences increase with $| \tilde \Lambda |$ 
(recall that the fragmentation in this region was 
very small). The energy difference between 
$\langle \phi_0^+ | H | \phi_0^+ \rangle $ and the 
exact ground state energy, $E_{\rm gs}$, relative to $E_{\rm gs}$ 
is shown in Fig.~\ref{fig:fig4} (a). This relative 
difference increases with $\Lambda$ and is zero only for 
$\Lambda=0$. Another measure of the capability of the 
mean-field state to describe the exact ground state 
is provided by the overlap of the trial state with 
the exact ground state. This overlap, 
$\langle \phi_{\rm gs} | \phi_0^+\rangle$ is plotted in 
Fig.~\ref{fig:fig4} (b). The overlap is 1 only 
for $\Lambda =0$. The differences in the spectral 
decomposition in the Fock space reflect in an overlap 
smaller than 1 when $\Lambda$ increases. Clearly at 
$\Lambda = -2$ the overlap between 
$\langle \phi_{\rm gs} | \phi_0^+\rangle$, decreases quickly 
and tends to zero for large values of $| \Lambda |$.

In the region $-3.2 < \tilde \Lambda < -2$, the minimum energy 
mean-field solutions, $| \phi_+^+\rangle$ and $| \phi_-^+\rangle$,
provide the same energy expectation value 
when the bias is not taken into account. 
The Fock decomposition of $| \phi_+^+\rangle$ is plotted 
in Fig.~\ref{fig:fig3}. The distribution 
for $| \phi_-^+\rangle$ would be symmetric 
with respect to $k=N/2$. The differences of the  
expectation energies, of these two states 
($| \phi_+^+\rangle$ and $| \phi_-^+\rangle$), 
with respect to the ground state energy are rather small, not only in 
the cat-like state region but also for larger values of 
$| \Lambda |$, where the difference tends to zero. 
On the contrary, the behavior of the overlap of these two 
states with the ground state is rather different. In 
the cat-state region, both overlaps are rather similar. 
The reason is that $| \phi_+^+ \rangle$ overlaps with 
the right part of the cat-state (in the Fock space) and 
$| \phi_-^+ \rangle$ overlaps with the left part of 
the cat-state. As $| \Lambda |$ is increased 
and the cat-state disappears, the presence of the bias 
term in the Hamiltonian ensures the breaking of the left/right 
symmetry by energetically promoting the $|L\rangle$ state.  
Thus, the system selects $| \phi_+^+ \rangle$ as the ground state, 
and therefore its overlap with the exact ground state 
tends to 1, while the overlap of 
$\langle \phi_{\rm gs} | \phi_-^+\rangle $ tends to zero.

The cat-state structure can be 
reproduced by defining as trial state the linear 
combination of $| \phi_+^+ \rangle$ and $| \phi_-^+ \rangle$, 
Eq.~(\ref{eq:cat}), as one can see by looking at the 
spectral decomposition of this state shown also in 
Fig.~\ref{fig:fig3}. Obviously, when $| \Lambda | $ 
increases, and the ground state is preferentially located 
in one of the wells, this state $| \phi_{\rm cat} \rangle $ 
does not give anymore a good reproduction of the Fock 
space decomposition of the ground state. If one looks 
at the overlap $\langle \phi_{\rm gs}| \phi_{\rm cat}\rangle $, 
this variational state clearly improves the overlap 
with the ground state in the cat-like region, but when 
$| \Lambda |$ increases, the overlap tends to a constant 
${1/\sqrt {2}}$. The behavior of the energy  can be 
observed in Fig.~\ref{fig:fig4} (a). We can see that 
there is an improvement in the cat-state region when 
using $| \phi_{\rm cat}\rangle$. However, 
when $| \Lambda |$ increases all three functions 
$| \phi_{\rm cat} \rangle$, $| \phi_+^+ \rangle$ 
and $| \phi_-^+ \rangle$ become degenerate in energy with the 
exact ground state.

\section{Improved global variational ansatz}
\label{crit}

We propose a variational ansatz that is valid independent of the 
strength of the interaction including at the same 
time the possibility of a mean-field and the 
existence of a cat-state. This state is 
a combination of two different mean-field states:
\beqa
|\Psi_{\rm var} \rangle &=&  A |\phi \rangle_1 + B |\phi \rangle_2 
= 
\frac {A}{{\sqrt {N!}}} 
\left [ \alpha a_L^{\dagger} + \beta a_R^{\dagger} \right ]^N| 0 \rangle 
\nonumber \\
&+& 
\frac {B}{{\sqrt {N!}}} 
\left [ \beta a_L^{\dagger} + \alpha a_R^{\dagger} \right ]^N 
| 0 \rangle  \, .
\label{eq:imp}
\eeqa
The variational parameters $\alpha$, $\beta$, $A$ and $B$ 
are taken real. Note that the two mean-field states 
are not necessarily orthogonal and therefore the 
normalization conditions are imposed in the following way:
\begin{equation}
\alpha^2 +\beta^2 = 1\,,\quad A^2+B^2+ 2(2 \alpha \beta)^N AB = 1 \, . 
\end{equation}
Let us discuss the differences of the ansatz in 
Eq.~(\ref{eq:imp}) with the states 
studied in the previous section. Here if $A$ 
or $B$ are zero, the state reduces to a mean-field 
state of the type considered before. 
On the other hand, if one constructs the 
combination of the two mean-field states and 
allows for a new minimization of the variational 
parameters, a noticeable improvement of 
the state is obtained. The expectation value of the 
Hamiltonian with this ansatz is given by
\beqa
\frac {E}{JN} &=&  \left[ - 2 \alpha \beta + \frac {\tilde \Lambda}{4} (1 -4 \alpha^2
\beta^2 ) + \frac {1}{2 \alpha \beta} \, \right] (A^2+B^2) \nonumber \\
&+& \frac {\varepsilon}{J} (\beta^2 -\alpha^2) (A^2-B^2)
+ \frac {\tilde \Lambda}{4} -\frac {1}{2 \alpha \beta}\,.
\label{eq:enermin} 
\eeqa

To determine the parameters of the variational state 
we follow two different criteria. The first consists in 
performing a numerical minimization of the expectation 
value of the energy, Eq.~(\ref{eq:enermin}). The many-body 
state thus computed is named 
$| \Psi_{\rm var}\rangle_{\rm min}$. In the second 
procedure, which can be pursued only when we already 
have a numerical solution of the exact ground 
state, we determine the coefficients by maximizing the 
overlap of the variational state with the exact 
ground state, giving the state $| \Psi_{\rm var}\rangle_{\rm max}$.

\begin{figure}[t]
\includegraphics[width=0.99\columnwidth, angle=0, clip=true]{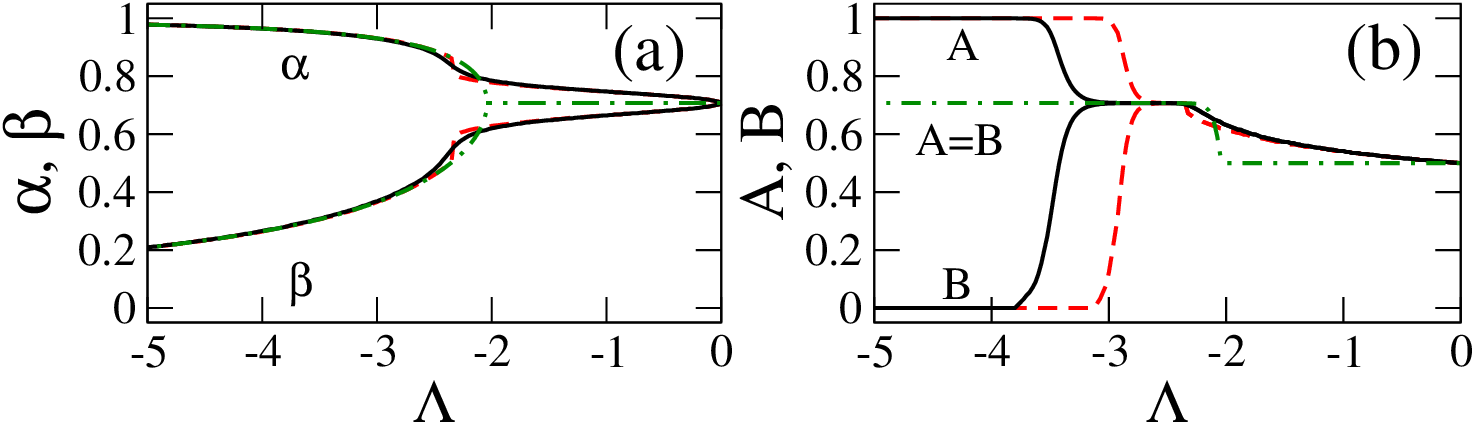}
\caption{
(Color online) Values of $\alpha$, $\beta$ (a) and of $A$, $B$ 
(b) obtained in the improved global variational approach of 
Eq.~(\ref{eq:imp}) by overlap maximization (black solid line) and 
by energy minimization (red dashed line), and the ones obtained 
in the variational approach of Eq.~(\ref{eq:cat}), for which $A=B$ 
(green dot-dashed line). 
\label{fig:fig5}}
\end{figure}

The first procedure, which does not require the previous 
numerical solution of the ground state, produces by construction 
the closest energy to the exact ground state energy within the form of 
Eq.~(\ref{eq:imp}). As will be discussed in the following, 
the second criteria although requiring the previous numerical 
solution of the ground state, produces in all cases an 
extremely close agreement with the ground state from the 
energetic point of view, while also improving the 
overlap with the numerically computed ground state. 
Thus, for certain applications where an analytical 
rendition of the state is preferable, our variational 
proposal\footnote{In the sense that $\langle \Psi_{\rm var}| H | \Psi_{\rm var} 
\rangle_{\rm max}$ provides also an upperbound to the ground state energy}
should be very useful.

The values of $\alpha,\beta$, $A$, and $B$ obtained in both 
cases are reported in Fig.~{\ref{fig:fig5}}. For 
$\Lambda=0$, we have $\alpha =\beta= 1/{\sqrt{2}}$ and 
$A=B=1/2$, recovering the function $| \phi_0^+\rangle$ 
that was the exact solution. This is the only case where 
our improved variational state coincides with 
the one proposed in Ref.~[\citenum{cirac98}]. Obviously in this case both 
conditions: minimum energy and maximum overlap, provide 
the same solution. Note that in this case, the overlap 
between the two components of the generalized variational 
ansatz ($| \phi\rangle_1$ and $| \phi\rangle_2$) is 
maximum, i.e. the two components coincide. 

When $|\Lambda|$ is increased, $-2<\Lambda<0$, $\alpha$ 
and $\beta$ become different while $A$ and $B$ remain equal but 
different from $1/2$. The improved variational state 
incorporates correlations beyond mean-field, 
and the overlaps $\langle \phi_{\rm gs} | \Psi_{\rm var}\rangle_{\rm min}$ 
and $\langle \phi_{\rm gs} | \Psi_{\rm var}\rangle_{\rm max}$, 
[see Fig.~\ref{fig:fig4} (b)] clearly improve with respect to 
$|\phi_0^+\rangle$. Also the difference in the 
expectation values of the energies 
$\langle \Psi_{\rm var} | H | \Psi_{\rm var}\rangle_{\rm min}$ 
and $\langle \Psi_{\rm var} | H | \Psi_{\rm var} \rangle_{\rm max}$ 
relative to the ground state energy become smaller as shown in 
Fig.~\ref{fig:fig4} (a).

\begin{figure}[t]
\includegraphics[width=0.8\columnwidth, clip=true]{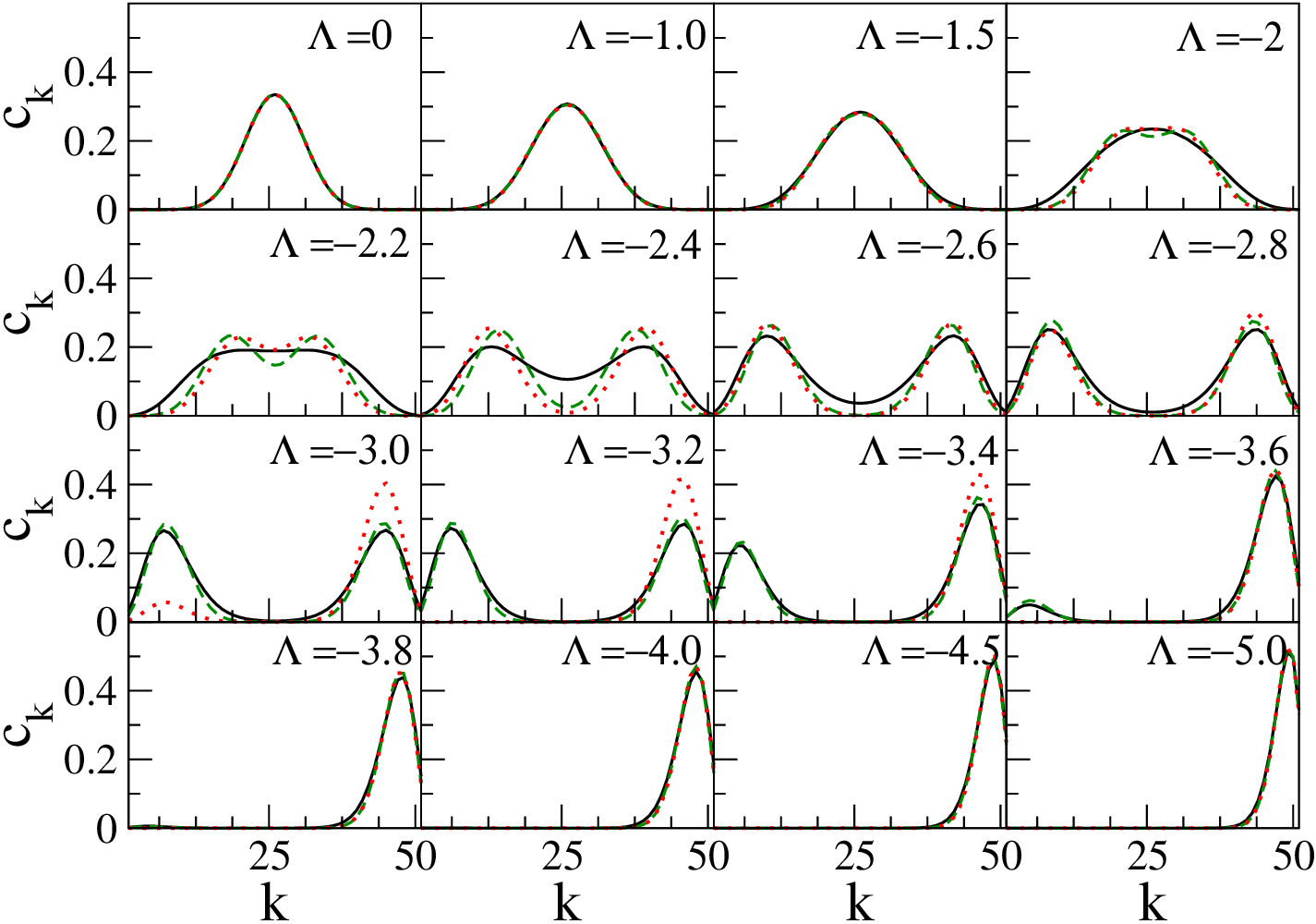}
\caption[]{(Color online) Spectral decomposition of the exact ground 
state (black solid line), and the states obtained using the 
ansatz defined in Eq.~(\ref{eq:imp}) when its overlap with 
the exact ground state is maximized (green dashed line) or when its energy 
is minimized (red dotted line). 
\label{fig:fig6}}
\end{figure}

In this region ($| \Lambda | < 2 $) the differences 
between the observables corresponding to these 
two variational states associated with the maximum 
overlap or with the minimum energy criteria are rather small. 
The state that minimizes the energy provides 
slightly better energies, however this difference is not 
significant in Fig.~\ref{fig:fig4} (a). Correspondingly the 
state that maximizes the overlap provides overlaps 
with the ground state closer to unity. However these 
differences are also not appreciable in Fig.~\ref{fig:fig4} (b). 
The Fock decomposition of these two variational states 
$| \Psi_{\rm var}\rangle_{\rm min}$ and 
$| \Psi_{\rm var}\rangle_{\rm max}$ compared with the one of 
the ground state are shown in Fig.~\ref{fig:fig6} for 
different values of $\Lambda$ and $N=50$. One can observe
a clear improvement of the Fock decomposition with respect 
to the mean-field state in this range of the 
interaction $| \Lambda |  < 2$.  

\begin{figure}[t]
\includegraphics[width=0.95\columnwidth,clip=true]{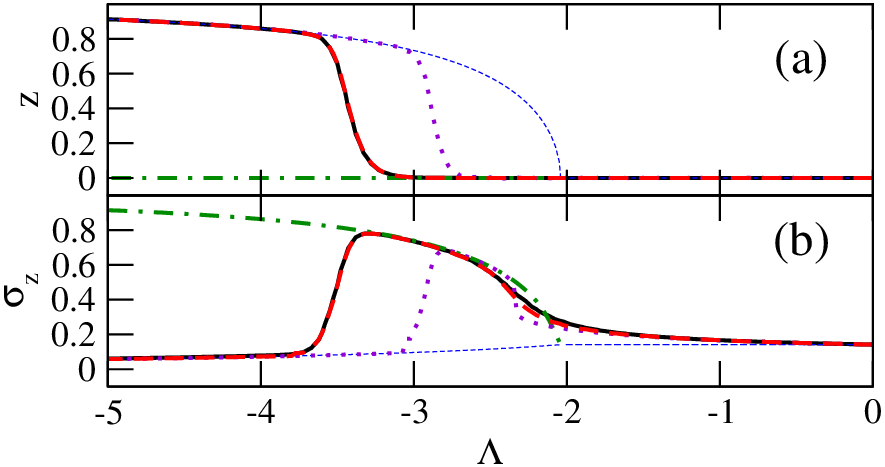}
\caption[]{(Color online) (a) Population imbalance $z$ 
and (b) its dispersion $\sigma_z$ as a function of $\Lambda$ for 
the exact calculation (black solid line), the improved global
variational approach in Eq.~(\ref{eq:imp}) by overlap maximization (red dashed line) and by energy minimization (violet dotted line)
The blue dashed line is the semi-classical prediction and the green 
dot-dashed line 
corresponds to the $| \phi_{\rm cat} \rangle_N$ of Eq.~(\ref{eq:cat}).
The number of particles is $N=50$.}\label{fig:fig7}
\end{figure}

Interestingly, the proposed state 
captures well the correlations beyond mean-field existing 
in the ground state of the problem before the classical 
bifurcation. These correlations, as discussed above 
and shown in Fig.~\ref{fig:fig2} (a), produce very small effects 
on the condensate fractions but become clearly visible when looking 
at the spectral decomposition of the ground state, see 
Fig.~\ref{fig:fig3}, or the dispersion of the population 
imbalance which is no longer corresponding to a 
simple binomial distribution, see Fig.~\ref{fig:fig2} (b). 

Once we cross the classical bifurcation, $\Lambda<-2$, the 
spectral decomposition of the ground state broadens and 
at $\Lambda\sim -2.2$ becomes quickly two-peaked. This 
region where the ground state has two maxima is what we 
refer to as the cat-state region. The main objective of 
the variational ansatz introduced in ~[\citenum{cirac98}], and discussed 
in Sect.~\ref{sec:bmf}, is to describe the ground state properties  
in this region. The results with the improved global 
variational ansatz of Eq.~(\ref{eq:imp}) are shown in 
Figs.~\ref{fig:fig4},~\ref{fig:fig6}, and~\ref{fig:fig7}. 
Unlike in the region before the bifurcation, here the two 
criteria used to compute the variational parameters provide 
fairly different results in some cases. The computed energy 
of the state is very close with both criteria, but its overlap with the ground 
state is different depending on the criteria used, 
see Fig.~\ref{fig:fig4}. This is a consequence of the clear
differences seen in the spectral decomposition, 
Fig.~\ref{fig:fig6}. The variational solution obtained by 
minimizing the energy is seen to depart from the exact 
solution in the region $-3.5 < \Lambda <-3$, predicting 
an earlier transition to the 'self-trapped' domain, 
see Figs.~\ref{fig:fig5} and~\ref{fig:fig7}. 

The criterion of maximizing the overlap implies the ansatz to follow much closer some 
of the explored ground state properties of the system as shown 
in Fig.~\ref{fig:fig7}, 
where the agreement with the exact calculation both for the 
population imbalance and its dispersion is extremely good 
in the considered domain. Thus, for these parameter values, 
obtaining a faithful representation of the ground state with the 
form proposed in Eq.~(\ref{eq:imp}) would require the 
prior numerical solution of the exact ground state.

\subsection{Fragmentation of the ground state}

\begin{figure}[t]
\includegraphics[width=0.95\columnwidth, angle=0, clip=true]{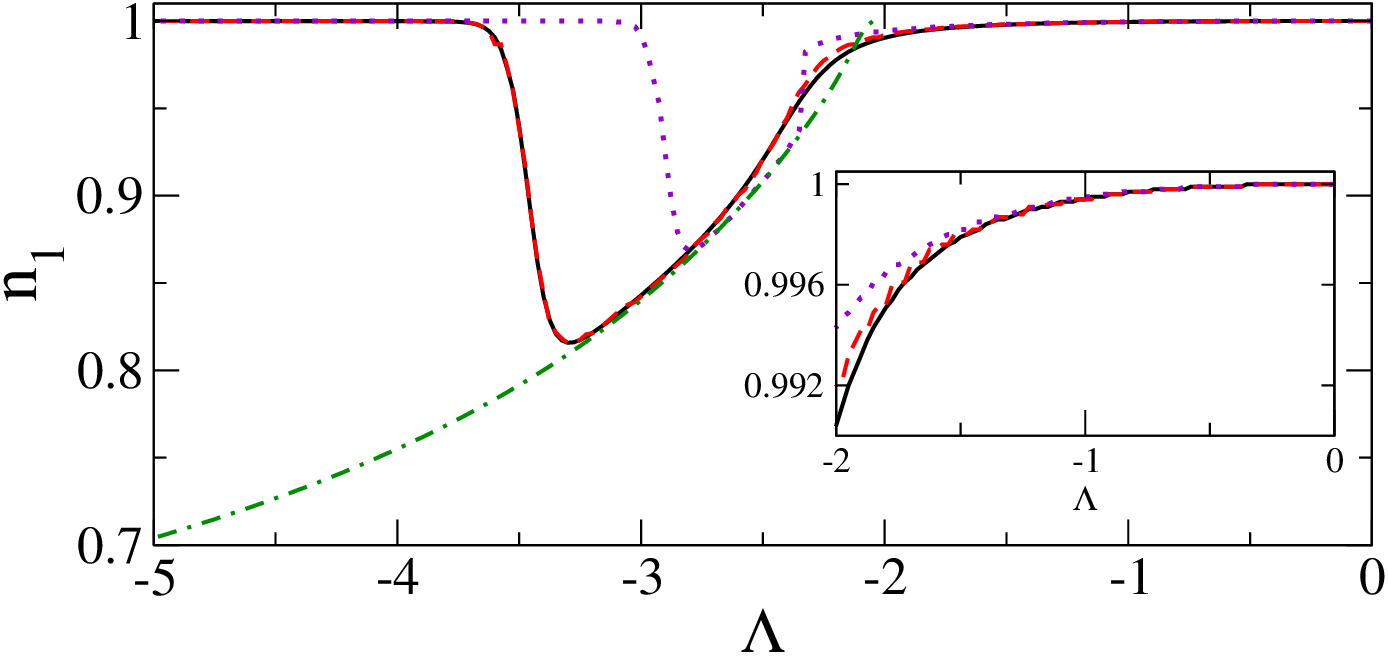}
\caption[]{(Color online) Largest eigenvalue of the one body density matrix, 
$n_1$, as a function of $\Lambda$ for the different many-body states 
discussed in the text: exact calculation (black solid line), 
$| \Phi_{\rm var}\rangle_{\rm max}$  (red dashed line), 
$| \Phi_{\rm var}\rangle_{\rm min}$  (violet dotted line), 
and the cat-state $| \phi_{\rm cat}\rangle $ 
of Ref.~[\citenum{cirac98}] (green dot-dashed line).
The inset shows the region before the bifurcation.} 
\label{fig:fig8}
\end{figure}

To complete the characterization of the proposed states we also 
study the fragmentation of the ground state of the system. To 
this end, we calculate the one-body density matrix and look 
at its larger eigenvalue. If the largest eigenvalue ($n_1$) is 
significantly smaller than unity, we have fragmentation and the system is not 
condensed in one single state. It also indicates the 
impossibility to describe the system by a mean-field state 
and therefore reveals the existence of correlations beyond 
mean-field. The largest eigenvalue of the one-body density 
matrix associated to the different states discussed 
in this work is reported in Fig.~\ref{fig:fig8} for different 
values of $| \Lambda |$. The two mean-field 
states, ($| \phi_0^+ \rangle, | \phi_+^+\rangle$), are not plotted as 
they have this eigenvalue equal to unity independently of 
$\Lambda$. The exact ground state gives rise to an $n_1$ very close 
to unity, in the region  $| \Lambda | < 2$. 
However, the eigenvalue is strictly one only for $\Lambda =0$, 
and is actually a smooth decreasing function of 
$| \Lambda |$. It decreases faster in the cat-like region 
reaching a minimum ($n_1 \sim 0.8 $) (maximal fragmentation) 
around $\Lambda =-3.2$. For larger values of $| \Lambda |$ 
it grows again reaching the value 1 as the system  
condenses in the left well due to the bias.

The $n_1$ associated to the variational state, 
$| \phi_{\rm cat}\rangle$, which exists for $| \Lambda | > 2$, 
starts from $n_1=1$ and decreases with increasing $| \Lambda |$ 
reproducing rather well the exact 
$n_1$. Contrary to what happens with the exact $n_1$, it continues 
decreasing and increasing the fragmentation failing to reproduce the region dominated 
by the bias. 
Finally, the variational many-body states 
proposed in the present paper, $| \Phi_{\rm var}\rangle_{\rm min}$ 
and $| \Phi_{\rm var}\rangle_{\rm max}$ reproduce very well 
the exact $n_1$ in the region before the bifurcation, where the 
system is slightly fragmented, see the inset in Fig.~\ref{fig:fig8}. 
This small fragmentation, as discussed above, indicates the 
presence of some correlations beyond the mean-field already in this region. 
In the cat-state region, the $| \Phi_{\rm var}\rangle_{\rm max}$ also  
reproduces the exact $n_1$, see
Fig.~\ref{fig:fig8}.

\section{Conclusions}
\label{sec4}

The variational analytical approach to the two-site Bose-Hubbard 
model gives a useful insight into the physical nature of the ground 
state of this apparently simple system that however shows a very 
rich phenomenology when the interaction or the number of particles 
change. We have carefully studied the limitations of the mean-field 
description strongly linked to the presence of fragmentation of 
the condensate and quantum fluctuations. The proposed variational 
state is able to describe rather well the exact state
and reproduces the energy, the one-body density matrix 
and thus the fragmentation of the state which are the main 
magnitudes that we have used to characterize the ground state. 

We have also compared the spectral decomposition of the exact ground 
state with the proposed state obtaining good agreement. 
The many-body states 
$| \Phi_{\rm var} \rangle_{\rm min}$, whose parameters 
are obtained by minimizing the energy, can be used for any 
number of particles. This state, incorporates for all $\Lambda$s, quantum 
correlations beyond the mean-field and reproduces very well 
the fragmentation induced by these correlations which become 
larger in the cat-state region.

The authors want to thank J. Martorell for a careful 
reading of the manuscript. M. M.-M. , is supported by 
an FPI grant from the MICINN (Spain). B.J.-D. is supported by a 
Grup Consolidat 2009SGR21 contract. This work is 
also supported by Grants No. FIS2008-01661, 
and No. 2009SGR1289 from Generalitat de Catalunya.


\end{document}